\documentclass[aps,prl,a4paper,twocolumn,groupedaddress,showpacs,nofootinbib]{revtex4}
\usepackage{amsmath}

\begin{document}

\title{Is the droplet theory for the Ising spin glass inconsistent
with replica field theory?
      }

\author{T.~Temesv\'ari}
\email{temtam@helios.elte.hu}
\affiliation{
Research Group for Theoretical Physics of the Hungarian Academy of Sciences,
E\"otv\"os University, P\'azm\'any P\'eter s\'et\'any 1/A,
H-1117 Budapest, Hungary}

\date{\today}

\begin{abstract}
Symmetry arguments are used to derive a set of exact identities between
irreducible vertex functions for the replica symmetric field theory of the
Ising spin glass in zero magnetic field. Their range of applicability
spans from mean field to short ranged systems in physical dimensions.
The replica symmetric theory is unstable for $d>8$, just like in
mean field theory. For $6<d<8$ and $d\lesssim 6$ the resummation of
an infinite number of terms is necessary to settle the problem.
When $d<8$, these Ward-like identities must be used to distinguish
an Almeida-Thouless line from the replica symmetric droplet
phase.
\end{abstract}


\maketitle

Field theory has proved to be an extremely useful tool in studying
critical transitions in ordinary systems, mostly by providing standard
methods like the loop expansion (above the upper critical dimension)
and the renormalization group \cite{Brezin_et_al}. Its adaptation to the
spin glass problem came just after the introduction of the Edwards--Anderson
model and application of the replica trick
\cite{EA}, resulting in a kind of replica field theory (see \cite{BrMo79}
and references therein). The first solution of the mean field theory
of the Ising spin glass 
provided a simple transition in zero external magnetic field from the
paramagnet to a replica symmetric (RS) spin glass state \cite{SK} which,
however, was later proved to be unstable \cite{AT}. This instability then
persisted perturbatively down to the upper critical dimension six \cite{BrMo79},
and even below it \cite{beyond}. Repair came soon, at least on the mean
field level, by the famous replica symmetry breaking (RSB) scheme of Parisi
(for details, see \cite{MePaVi}): the spin glass transition is now to a
lower symmetry phase which is marginally stable for all $T<T_c$, i.e.\ %
it is a massless phase, and has a special --- ultrametrical --- hierarchy.
The RSB transition has also the peculiarity of extending to nonzero magnetic
fields along the Almeida--Thouless (AT) transition {\em line\/} \cite{AT}.

From this point on, the spin glass community has become highly divided
about the type of the transition and the structure of the spin glass
state of short ranged, finite-dimensional models. Supporters of the RSB
scenario followed the classical route trying to build a field theory on
the basis of the --- highly nontrivial --- Parisi solution \cite{beyond}.
As it turned out from the physical interpretation of RSB \cite{MePaVi},
the Parisi theory corresponds to a complicated ergodicity breaking
with a Gibbs state decomposed into many pure states. On the other hand,
the so called droplet theory \cite{BrMo86,FiHu86}, which was developed
mostly on the original lattice system, has a simpler phase structure
with only two pure states which are related by the spin inversion
symmetry (just like in a ferromagnet), and it predicts that a magnetic
field destroys the transition. In the droplet picture the mode
called replicon (R) remains massless in the whole spin glass phase
\cite{BrMo86}
providing the only common feature both theories share.

The simple phase structure of the droplet theory implies an RS phase
with a nonzero order parameter, the corresponding replica field theory
has a Lagrangean which is invariant for any permutation of the $n$
replicas (up to cubic order, it was displayed in Ref.~\cite{rscikk}).
Since the coupling constants of such a field theory are chosen by symmetry,
or --- alternatively --- thought to be the outcome of summing out short
ranged fluctuations down to the momentum cutoff $\Lambda$, they have little
memory of the original control parameters like temperature and magnetic
field. As a nonzero magnetic field does not change the symmetry, it is
difficult to distinguish between a zero-field RS spin glass phase and an
AT line (massless and replica symmetric too). In this letter, we use exact
symmetry arguments to find out the thermal route of the spin glass in
zero magnetic field when crossing $T_c$ from the paramagnet having the
Lagrangean
\begin{widetext}
\begin{align}\label{Lagrangean}
&\mathcal{L}=\frac{1}{2}\sum_{\mathbf p}
\Big(\frac{1}{2} p^2+\bar{m}_1\Big)\sum_{\alpha\beta}
\phi^{\alpha\beta}_{\mathbf p}\phi^{\alpha\beta}_{-\mathbf p}
-\frac{1}{6\sqrt{N}}
\sideset{}{'}\sum_{\mathbf {p_i}}
\bar{w}_1\sum_{\alpha\beta\gamma}\phi^{\alpha\beta}
_{\mathbf p_1}\phi^{\beta\gamma}_{\mathbf p_2}
\phi^{\gamma\alpha}_{\mathbf p_3}\\
&-\frac{1}{24 N} \sideset{}{'}\sum_{\mathbf {p_i}}
\Big( \bar{u}_1\sum_{\alpha\beta\gamma\delta}\phi^{\alpha\beta}
_{\mathbf p_1}\phi^{\beta\gamma}_{\mathbf p_2}
\phi^{\gamma\delta}_{\mathbf p_3}\phi^{\delta\alpha}_{\mathbf p_4}+
\bar{u}_2\sum_{\alpha\beta}\phi^{\alpha\beta}
_{\mathbf p_1}\phi^{\alpha\beta}_{\mathbf p_2}
\phi^{\alpha\beta}_{\mathbf p_3}\phi^{\alpha\beta}_{\mathbf p_4}+
\bar{u}_3\sum_{\alpha\beta\gamma}\phi^{\alpha\gamma}
_{\mathbf p_1}\phi^{\alpha\gamma}_{\mathbf p_2}
\phi^{\beta\gamma}_{\mathbf p_3}\phi^{\beta\gamma}_{\mathbf p_4}+
\bar{u}_4\sum_{\alpha\beta\gamma\delta}\phi^{\alpha\beta}_{\mathbf p_1}
\phi^{\alpha\beta}_{\mathbf p_2}\phi^{\gamma\delta}_{\mathbf p_3}
\phi^{\gamma\delta}_{\mathbf p_4}\Big)+\dots,\notag
\end{align}
\end{widetext}
where the classical fields $\phi^{\alpha\beta}$ are symmetrical in the
replica indices $\alpha$, $\beta =1,\dots,n$ with zero diagonal, and
momentum conservation is understood in the primed sums. The number of spins
$N$ goes to infinity in the thermodynamic limit, while $n\to 0$ ensures
the spin glass limit. (A barred notation for the {\em bare\/} coupling
constants is used to distinguish them from the corresponding {\em exact\/}
vertex functions.)
In addition to the permutational symmetry of the
presumed low temperature phase, the invariants in (\ref{Lagrangean})
have the extra attribute that a replica occurs
always an even number of times; a Hubbard--Stratonovich-like derivation
\cite{BrMo79,rscikk}
of (\ref{Lagrangean}) makes this point evident. In mathematical form,
$\mathcal L$ is invariant under the transformation ${\phi'}^{\alpha\beta}
=(-1)^{\alpha+\beta}\,\phi^{\alpha\beta}$.
Following that transformation by the special permutation of grouping odd
and even replicas separately --- i.e.\ $(1,2,3,\dots,n)\longmapsto
(1,3,5,\dots,2,4,6,\dots)$ --- may lead us to figure out that
we can define
a class of transformations
leaving $\mathcal L$ of (\ref{Lagrangean}) invariant
as follows: Let us divide the $n$ replicas
into two groups consisting of $p$ and $n-p$ elements (p being a free
parameter). For the transformed field $\phi'$ we have:
\begin{equation}\label{O} {\phi'}^{\alpha\beta}=
\begin{cases}
\phi^{\alpha\beta} & \text{for $\alpha$ and $\beta$ in the same group,}\\
-\phi^{\alpha\beta} & \text{for $\alpha$ and $\beta$ in different groups.}
\end{cases}
\end{equation}
Therefore the paramagnetic phase has a higher symmetry than even the simplest,
generic replica symmetric, spin glass phase, and the presumed paramagnet to %
droplet spin glass transition breaks that higher symmetry in replica field
theory.

Proceeding further, we can follow the steps of Ref.~\cite{reparametrization}
to conclude {\em exact\/} identities between the irreducible vertices of
the low temperature RS phase. Including a source term
$- \sum_{\alpha\beta} h_{\alpha\beta}\phi^{\alpha\beta}_{\mathbf p=0}$ into
(\ref{Lagrangean}), a Legendre-transformed free energy  $F(\mathbf q)$ can be
derived, and it is invariant under the obviously orthogonal transformation
$\mathbf O$
of Eq.~(\ref{O}) \footnote{The $n(n-1)/2$ $q_{\alpha\beta}$ can be arranged into
the vector $\mathbf q$, and the transformation in Eq.~(\ref{O}) may be written
then as $\mathbf q'=\mathbf O \mathbf q$ with the diagonal tranformation
matrix $\mathbf O$ having the properties $O_{\alpha\beta,\alpha\beta}=1$
or $-1$, depending on $\alpha$ and $\beta$ belonging to the same group or
not.}. The derivatives of $F$ provide the
zero-momentum one-particle irreducible vertices
\cite{Brezin_et_al}; their definitions are the following:
\begin{align}
-H_{\alpha\beta} &=\frac{\partial F}{\partial q_{\alpha\beta}}\,\,,  &
M_{\alpha\beta,\gamma\delta} &=\frac{\partial^2 F}{\partial q_{\alpha\beta}
\partial q_{\gamma\delta}}\,\,, \notag
\end{align}
and similar formulae for $W_{\alpha\beta,\gamma\delta,\mu\nu}$, %
$U_{\alpha\beta,\gamma\delta,\mu\nu,\rho\omega}$, \ldots\ etc.
From $F(\mathbf q')=F(\mathbf q)$ and the orthogonality of $\mathbf O$
follows (using tensorial notation):
\begin{equation}\label{trafo}
\mathbf{H'}=\mathbf{O}\mathbf{H}\,,\qquad\mathbf{M'}=\mathbf{O}\mathbf{M}
\mathbf{O}^{-1}\,,
\end{equation}
and analogous relations for higher order terms. As $p$ is a free parameter,
$\mathbf O$ is in fact a continuous symmetry transformation, and assuming
that $\mathbf q$ is replica symmetric, it is easy to derive $\sqrt{(
\mathbf q'-\mathbf q)^2}=2q\sqrt{p(n-p)}$, providing an infinitesimal
transformation for $p(n-p)$ small. We can, therefore, expand the left hand
sides of Eq.~(\ref{trafo}) around $\mathbf q$, and equating the coefficients
of identical powers of $p$ and $n-p$ on both sides. Assuming stationarity,
i.e.\ $\mathbf{H}=0$, and remembering that we are in a RS state with a
nonzero order parameter $q$, several identities can be derived in this
way between the vertices $m_1$, $m_2$, $m_3$; $w_1$, \ldots, $w_8$;
$u_1$, \ldots etc.\ \footnote{This identities are much simpler when displayed
in terms of the set of vertices with the lower case notation
(their bare counterparts are the coupling constants in front of the
invariants with the unrestricted sums in the Lagrangean).
The linear
relationship between these two sets of parameters were derived in
\cite{rscikk} for the $\bar m$'s and $\bar w$'s; Eqs.\ (20)--(24) and Table~1.
As a property of the generic RS symmetry, it is not restricted to the
bare couplings, but is valid for the exact vertices too. The formulae for the
23 quartic couplings are more complicated, and will be detailed in a longer
publication.}. These are,
however, --- unlike traditional Ward-identities --- 
power series of $q$ with higher and higher order vertices, the most
prominent ones are displayed here:
\begin{align}
m_1+\frac{n}{2}m_2 &=-\big(w_2+\frac{n}{2}w_3+\frac{n^2}{2}w_6\big)
\,q\notag\\\label{m1}
&\mathrel{\phantom{=}}+\frac{2}{3}
(u_2+\dots)\,q^2+\dots\,, \\[3pt]\label{m2}
m_2 &=-\Big[\big(w_1+\frac{1}{3}w_3\big)+n\,\big(\frac{1}{3}w_5+w_6\big)\Big
]\,q+\dots\,, \\ \label{m3}
m_3 &=-\Big[\big(w_4+\frac{1}{2}w_5-\frac{1}{2}w_6\big)+\frac{n}{2}w_7
\Big]\,q+\dots\,;
\end{align}
following from the first and
\begin{align}\label{w2}
\big(w_2+\frac{n}{2}w_3+\frac{n^2}{2}w_6\big) &=
(u_2+\dots)\,q+\dots 
\end{align}
from the second equation in Eq.~(\ref{trafo}). (Ellipsis dots in the above
expressions are to substitute higher order terms in $n$ or $q$.)

The value of the identities in Eqs.~(\ref{m1})--(\ref{w2}),
and the others not displayed here, rests on their generality --- their
derivation used only symmetry arguments ---; as a result, they must be valid
for the mean field as well as for low dimensional systems. It is tempting
to solve these equations iteratively, i.e.\ assuming all the vertices are
analytical in $q$. The most important result we can get in this way is
the famous instability \cite{AT} of the replicon mass $\Gamma_{\text R}$
for $n\to 0$ and $u_2>0$ at criticality:
\begin{equation}\label{R}
\Gamma_{\text R}=2m_1= -\frac{2}{3} u_2\, q^2+O(q^3)\,.
\end{equation}
Moreover, all the vertices incompatible with the symmetry of the paramagnetic
phase are expressible in terms of those present for $T>T_c$ too. In leading
order in $q$ we have $m_2=-w_1q$, $w_2=u_2q$, $w_3=u_3q$, $w_4=u_4q$ and
$w_5=u_1q$, while all the others ($m_3$, $w_6$, $w_7$, $w_8$) are of order
$q^2$. All these results can be verified for mean field theory (for a
generic $n$) using the explicit formulae of Ref.~\cite{rscikk}
and their extensions to the quartic order, and exploiting
the fact that bare and exact parameters are identical for a zero loop
calculation.

We cannot follow this procedure in {\em any\/} finite dimension $d$, as the
exact vertices are no longer analytical as a function of $q$. Nevertheless,
Eqs.~(\ref{m1})--(\ref{w2}) can now be used perturbatively, for $d>6$, to
compute the {\em bare\/} parameters as a function of $q$, and they will have,
beside the analytical part, terms with noninteger, $d$-dependent powers.
Instead of Eq.~(\ref{R}), we now have for
the deviation of the bare replicon mass from its critical value:
\begin{equation}\label{bareR}
2(\bar{m}_1-\bar{m}_{1c})= -\frac{2}{3} \bar{u}_2\, q^2+C_d\,\bar{w}_1^2\,(\bar{w}_1q)^
{1-\frac{\epsilon}{2}}\,,
\end{equation}
where $\epsilon =6-d$. $C_d$ --- unlike the coefficient of the quadratic
term --- can {\em not\/} be computed by a simple truncation of
(\ref{m1}), since a contribution from an arbitrary $k$-point vertex
(multiplied by $q^{k-2}$) must be included \footnote{The one-loop
$k$-leg ``bubble" gives a $(\bar{w}_1q)^{3-k-\frac{\epsilon}{2}}$
contribution.}. 

At that point, two important remarks are appropriate. Firstly, there is
some ambiguity in assigning the bare coupling constants to a given physical
state below $T_c$ fixed by $q$: an offset of the zero-momentum fields,
$\phi^{\alpha\beta}_{\mathbf p=0}\longrightarrow
\phi^{\alpha\beta}_{\mathbf p=0}-\sqrt N\,\Phi$, leaves all the irreducible
vertices unaltered while the bare couplings changing. We can get rid of
``tadpole" insertions by the choice
$\Phi=q$ rendering the one-point function zero \cite{BrMo79}. We use that
case for a unique definition of the bare parameter space throughout this
letter. Secondly, we must emphasize that perturbation theory remains
valid for $d>6$ in the loop-expansion sense, since $C_d$ as well as the
coefficient of the $q^2$ term in (\ref{bareR}), and in fact any quantity,
can be computed,
at least in principle, in terms of $\bar{w}_1^2$, $\bar{u}_1$,\dots,$\bar{u}_4$,
etc. Nevertheless, unlike the analytical contributions which can be computed
by the truncation method just as in the mean field case, to get the
$d$-dependent powers at a given order of the loop-expansion, we must resum
an infinite number of terms (all the ``bubbles" for $C_d$ at 1-loop
order). (We are not able to do this at the moment, though it may not be
a completely hopeless task.) 

An evaluation of Eq.~(\ref{m2}) at one-loop level provides us
\begin{equation}\label{bareRA}
\bar{m}_2=-\bar{w}_1 q\,[1+O(\bar{w}_1^4)]+
C'_d\,\bar{w}_1^2\,(\bar{w}_1q)^
{1-\frac{\epsilon}{2}}\,,
\end{equation}
with $C'_d$ again comprising an infinite number of terms, and there is
a good chance that the linear term is exactly $-\bar{w}_1 q$, i.e.\
it originates from the offset of the fields alone.

After elucidating the
$q$-dependence of the bare parameters, expressions for the exact masses can
be straightforwardly computed at the one-loop level:
\begin{align}\label{exactR}
\Gamma_{\text{R}}&= -\frac{2}{3} \bar{u}_2\, q^2\\
&\mathrel{\phantom=}- 16\bar{w}_1^2
(\bar{w}_1q)^
{1-\frac{\epsilon}{2}} \left[\,\int \frac{d^dp}{(2\pi)^d} \frac{1}{p^4
(p^2+2)^2}+\dots \,\right]\notag
\end{align}
for $n=0$ and (keeping a generic $n$ now)
\begin{equation}\label{exactRA}
\Gamma_{\text{A}}-\Gamma_{\text{R}}=(n-2)m_2=-(n-2)\bar{w}_1q
\left(1-\frac{n-2}{\epsilon}\bar{w}_1^2\right)\,.
\end{equation}
The dots in (\ref{exactR}) refer to the infinite terms coming from
the 5-leg, 6-leg, etc.\ vertices. (Similar terms in (\ref{exactRA}) are not
displayed, as they are definitely subleading when $d>6$.) While the anomalous
mode (A) behaves regularly for any $d>6$ and is massive, there is a competition
between the two terms in $\Gamma_{\text{R}}$: as long as $d>8$ the
$d$-dependent power in (\ref{exactR}) is subleading, and the truncation method
to calculate leading terms works as well as for mean field theory. We thus
conclude that the RS phase is unstable, and RSB characterizes the spin glass
phase for $d>8$. 

Stability depends on the infinite sum in Eq.~(\ref{exactR}), which is not
available at the moment, when $6<d<8$.
We are taking the opportunity
to comment on earlier works now. In Ref.~\cite{BrMo79}, the second term for
$\Gamma_{\text{R}}$ without the infinite sum represented by the dots was
obtained, and --- as it is manifestly negative ---
it was inferred that instability thus persisted
below 8 dimensions. The authors, however, found a sophisticated way to
construct a {\em different\/} RS solution which is marginally stable,
i.e.\ $\Gamma_{\text{R}}=0$. (A modified version of that solution was
proposed  in a recent paper as a candidate model for the droplet theory
\cite{Cirano}.) The first result was based on the traditional way to
build up a symmetry-broken theory from the symmetrical Lagrangean
(\ref{Lagrangean}): the bare parameters are continued analytically
below $T_c$, and the offset of the fields, as explained above, then
breaks the symmetry. This procedure is obviously correct for the leading
terms, e.g.\ the $O(q)$ term of (\ref{bareRA}) is reproduced, and it gives
definite predictions for the coefficients $C_d$, $C'_d$, \ldots etc.
These, however, should be justified by showing that they satisfy the
exact identities above. As a matter of fact, the second procedure of
Ref.~\cite{BrMo79} resulting in
the massless replicon mode is also correct in leading order; not surprisingly
$C_d$ is tuned to shift $\Gamma_{\text{R}}$ to zero. The most remarkable
observation concerns the inclusion of an external magnetic field
$H_{\text{ext}}$
which scales as $H_{\text{ext}}^2\sim q^{2-\frac{\epsilon}{2}}$.
It contributes also
to $C_d$, and may render the replicon mode massless \cite{GrMoBr83}.
The only way to distinguish an AT line from a dropletlike phase is by means of
the exact identities which exploit the extra symmetry of the zero-field case.

Finally we turn now to the $d<6$ case, where the loop-expansion breaks down,
and the perturbative renormalization group elaborated in \cite{Iveta} takes
over its role. What follows from now on is highly based on the details
of Ref.~\cite{Iveta}. To study the crossover region around the zero-field
fixed point $\bar{w}_1^{*2}=-\epsilon/(n-2)$, it is useful to introduce
Wegner's nonlinear scaling fields \cite{Wegner} defined
by the exact renormalization equations $\dot{g_i}=\lambda_i g_i$ where
$\lambda_i$'s are the scaling exponents. The bare parameters can be expressed
in terms of the $g_i$'s, as displayed here for the three masses
(omitting nonlinear terms and irelevant fields):
\begin{align*}
\bar{m}_1&=\frac{n-2}{4}\bar{w}_1^{*2}+(g_1+2g_2+g_3)+\dots\,,\\
\bar{m}_2&=-(g_2+g_3)+\dots\,,\\
\bar{m}_3&=\frac{1}{4}g_3+\dots\,\,\,.
\end{align*}
Assuming the form $g_i\cong C_i
(\bar{w}_1^*q)^{z_i}$, 
we can conclude from (\ref{m3}) that
$C_3=O(\epsilon)$, hence hindering us to compute the correction term of
$z_3=1+\dots$ at that order. The other two exponents, however, can be
derived from the logarithms of (\ref{m1}) and (\ref{m2}) providing
$z_1=1-\epsilon/2+O(\epsilon^2)$ and $z_2=1+O(\epsilon^2)$.
These results are in accord with the relations $\lambda_i=
(2-\epsilon/2+\eta/2)\,z_i$  --- which must
follow from the flow of the order parameter $\dot{q}\cong
(\beta/\nu)\,q$, with $\beta/\nu=
(2-\epsilon/2+\eta/2)$ --- when comparing with the independent
calculations of the $\lambda_i$'s and $\eta$
in \cite{Iveta}. Although the leading
terms for the $C_i$'s are trivial [$C_1=(n-2)+O(\epsilon)$, $C_2=1
+O(\epsilon)$ and $C_3=O(\epsilon)$], the $O(\epsilon)$ corrections
can be obtained again, just like for $6<d<8$, including all the terms of
Eqs.~(\ref{m1}--\ref{w2}). To see this, let us consider the scaling form
of a generic $k$-point vertex 
\begin{equation*}
\Gamma^{(k)}={|g_1|}^{\frac{1}{\lambda_1}[d-k(2-\frac{\epsilon}{2}+\frac{\eta}{2})]}
\,\,\tilde{\Gamma}^{(k)}(x,y)\,,
\end{equation*}
where we have used the common notation $\Gamma$ for the vertices (instead of
the $m$'s, $w$'s, $u$'s, etc). The two scaling variables are
$x=g_2/{|g_1|}^{\frac{\lambda_2}{\lambda_1}}=g_2/{|g_1|}^{\frac{z_2}{z_1}}$ and $y=
g_3/{|g_1|}^{\frac{\lambda_3}{\lambda_1}}=g_3/{|g_1|}^{\frac{z_3}{z_1}}$, whereas
the $\tilde{\Gamma}^{(k)}$'s are scaling functions
specific to the given vertex from the $k$-point family.
As we have for the leading terms in $\epsilon$:
$\tilde{\Gamma}^{(k)}\sim \bar{w}_1^{*k}$
($k\ge 3$ and forgetting now about the only exception $w_1$), a typical term
of, say, Eq.~(\ref{m1}):
\begin{equation*}
q^{k-2}\,\Gamma^{(k)}\sim
q^{k-2}\,
(\bar{w}_1^*q)^{(d\frac{\nu}{\beta}-k)}\,\bar{w}_1^{*k}
\sim \bar{w}_1^{*2}\,(\bar{w}_1^*q)^{\frac{\gamma}{\beta}}\,;
\end{equation*}
i.e.\ all terms are the same $\epsilon$ order.

To test stability below six dimensions, $\tilde{\Gamma}_R(x,y)$ must be
computed in $\epsilon$-expansion first, and then substituting
$x=C_2/|C_1|^{\frac{z_2}{z_1}}$, $y=C_3/|C_1|^{\frac{z_3}{z_1}}$
provides us the replicon mass for $T\lesssim T_c$ in zero magnetic field.
From $x=1/(2-n)+O(\epsilon)$ and $y=O(\epsilon)$ we can conclude:
\[
\tilde{\Gamma}_{\text{R}}=(-1+2x+y)+\dots=\frac{n}{2-n}+O(\epsilon).
\]
In the spin glass limit $n\to 0$, stability depends on the correction
term which contains two contributions: the first one comes from the correction
of the scaling function $\tilde{\Gamma}_{\text{R}}$
--- which has been computed and gives a negative (unstable) result ---,
and the second one from those of $x$ and $y$. These are, however,
--- as argued above --- not available at the moment, as they result from
a resummation of an infinite number of terms in the identity of Eq.~%
(\ref{m1}).

To conclude, the importance to find the correct trajectory in the bare
parameter space in absence of an external magnetic field
(what is called here the thermal route) is emphasized when the stability of
the low temperature RS phase --- the droplet phase --- is tested. We argued
that for $6<d<8$ and $d\lesssim 6$ an infinite number of one-loop graphs
should be resummed to settle the problem. Without using the Ward-like
identities derived in this letter, the droplet phase cannot be distinguished
from an AT line for $d<8$.
Our conclusions are in conflict with those of a
recent paper by M.~A.~Moore \cite{Moore}. Although both of us realized the
necessity to resum an infinite number of terms, in Ref.~\cite{Moore}
self-energy graphs with different {\em loops\/} are proposed to be summed
to get rid of infrared divergences caused by the replicon mode. In this
letter, we argue that allowing for all the one-loop graphs
with different {\em number of legs\/} is a must to ensure the extra symmetry
imposed by the lack of an external magnetic field.
Old beliefs that replica field theory is inconsistent with the droplet
picture were based on the instability emerging at one-loop order if
Eq.~(\ref{exactR}) without the infinite terms is used, which we believe
is not correct.
Therefore the conclusion that the RS phase is unstable for $6<d<8$
is thus premature,
just as that it is stable for $d<6$.
Eq.~(\ref{exactR}) loses its relevance for $d<6$,
and the study of the crossover region around the nontrivial fixed point is
inevitable. It must be stressed that our identities are valid even in
3 dimensions, though perturbative methods are not
available then to take advantage of them.
\begin{acknowledgments}
Discussions with C.~De Dominicis are highly appreciated. I am grateful to
M.~A.~Moore for sending me comments to the e-print version of that letter.
\end{acknowledgments}


\begin{thebibliography}{16}
\expandafter\ifx\csname natexlab\endcsname\relax\def\natexlab#1{#1}\fi
\expandafter\ifx\csname bibnamefont\endcsname\relax
  \def\bibnamefont#1{#1}\fi
\expandafter\ifx\csname bibfnamefont\endcsname\relax
  \def\bibfnamefont#1{#1}\fi
\expandafter\ifx\csname citenamefont\endcsname\relax
  \def\citenamefont#1{#1}\fi
\expandafter\ifx\csname url\endcsname\relax
  \def\url#1{\texttt{#1}}\fi
\expandafter\ifx\csname urlprefix\endcsname\relax\def\urlprefix{URL }\fi
\providecommand{\bibinfo}[2]{#2}
\providecommand{\eprint}[2][]{\url{#2}}

\bibitem[{\citenamefont{Br{\'e}zin et~al.}(1976)\citenamefont{Br{\'e}zin,
  Le~Guillou, and Zinn-Justin}}]{Brezin_et_al}
\bibinfo{author}{\bibfnamefont{E.}~\bibnamefont{Br{\'e}zin}},
  \bibinfo{author}{\bibfnamefont{J.~C.} \bibnamefont{Le~Guillou}},
  \bibnamefont{and}
  \bibinfo{author}{\bibfnamefont{J.}~\bibnamefont{Zinn-Justin}},
  \emph{\bibinfo{title}{Field Theoretical Approach to Critical Phenomena}}
  (\bibinfo{publisher}{Academic Press}, \bibinfo{year}{1976}),
  vol.~\bibinfo{volume}{6} of \emph{\bibinfo{series}{Phase Transitions and
  Critical Phenomena}}, chap.~\bibinfo{chapter}{3}.

\bibitem[{\citenamefont{Edwards and Anderson}(1975)}]{EA}
\bibinfo{author}{\bibfnamefont{S.~F.} \bibnamefont{Edwards}} \bibnamefont{and}
  \bibinfo{author}{\bibfnamefont{P.~W.} \bibnamefont{Anderson}},
  \bibinfo{journal}{J. Phys. F} \textbf{\bibinfo{volume}{5}},
  \bibinfo{pages}{965} (\bibinfo{year}{1975}).

\bibitem[{\citenamefont{Bray and Moore}(1979)}]{BrMo79}
\bibinfo{author}{\bibfnamefont{A.~J.} \bibnamefont{Bray}} \bibnamefont{and}
  \bibinfo{author}{\bibfnamefont{M.~A.} \bibnamefont{Moore}},
  \bibinfo{journal}{J. Phys. C} \textbf{\bibinfo{volume}{12}},
  \bibinfo{pages}{79} (\bibinfo{year}{1979}).

\bibitem[{\citenamefont{Sherrington and Kirkpatrick}(1975)}]{SK}
\bibinfo{author}{\bibfnamefont{D.}~\bibnamefont{Sherrington}} \bibnamefont{and}
  \bibinfo{author}{\bibfnamefont{S.}~\bibnamefont{Kirkpatrick}},
  \bibinfo{journal}{\prl} \textbf{\bibinfo{volume}{35}}, \bibinfo{pages}{1792}
  (\bibinfo{year}{1975}).

\bibitem[{\citenamefont{de~Almeida and Thouless}(1978)}]{AT}
\bibinfo{author}{\bibfnamefont{J.~R.~L.} \bibnamefont{de~Almeida}}
  \bibnamefont{and} \bibinfo{author}{\bibfnamefont{D.~J.}
  \bibnamefont{Thouless}}, \bibinfo{journal}{J. Phys. A}
  \textbf{\bibinfo{volume}{11}}, \bibinfo{pages}{983} (\bibinfo{year}{1978}).

\bibitem[{\citenamefont{De~Dominicis et~al.}(1998)\citenamefont{De~Dominicis,
  Kondor, and Temesv\'ari}}]{beyond}
\bibinfo{author}{\bibfnamefont{C.}~\bibnamefont{De~Dominicis}},
  \bibinfo{author}{\bibfnamefont{I.}~\bibnamefont{Kondor}}, \bibnamefont{and}
  \bibinfo{author}{\bibfnamefont{T.}~\bibnamefont{Temesv\'ari}},
  \emph{\bibinfo{title}{Beyond the Sherrington-Kirkpatrick Model}}
  (\bibinfo{publisher}{World Scientific}, \bibinfo{year}{1998}),
  vol.~\bibinfo{volume}{12} of \emph{\bibinfo{series}{Series on Directions in
  Condensed Matter Physics}}, p. \bibinfo{pages}{119},
  \eprint{cond-mat/9705215}.

\bibitem[{\citenamefont{M{\'e}zard et~al.}(1987)\citenamefont{M{\'e}zard,
  Parisi, and Virasoro}}]{MePaVi}
\bibinfo{author}{\bibfnamefont{M.}~\bibnamefont{M{\'e}zard}},
  \bibinfo{author}{\bibfnamefont{G.}~\bibnamefont{Parisi}}, \bibnamefont{and}
  \bibinfo{author}{\bibfnamefont{M.~A.} \bibnamefont{Virasoro}},
  \emph{\bibinfo{title}{Spin glass theory and beyond}},
  vol.~\bibinfo{volume}{9} of \emph{\bibinfo{series}{Lecture Notes in Physics}}
  (\bibinfo{publisher}{World Scientific}, \bibinfo{address}{Singapore},
  \bibinfo{year}{1987}).

\bibitem[{\citenamefont{Bray and Moore}(1986)}]{BrMo86}
\bibinfo{author}{\bibfnamefont{A.~J.} \bibnamefont{Bray}} \bibnamefont{and}
  \bibinfo{author}{\bibfnamefont{M.~A.} \bibnamefont{Moore}}, in
  \emph{\bibinfo{booktitle}{Proceedings of the Heidelberg Colloquium on Glassy
  Dynamics}}, edited by \bibinfo{editor}{\bibfnamefont{J.~L.} \bibnamefont{van
  Hemmen}} \bibnamefont{and}
  \bibinfo{editor}{\bibfnamefont{I.}~\bibnamefont{Morgenstern}}
  (\bibinfo{publisher}{Springer}, \bibinfo{year}{1986}), vol.
  \bibinfo{volume}{275} of \emph{\bibinfo{series}{Lecture Notes in Physics}},
  \bibinfo{note}{and references therein}.

\bibitem[{\citenamefont{Fisher and Huse}(1986)}]{FiHu86}
\bibinfo{author}{\bibfnamefont{D.~S.} \bibnamefont{Fisher}} \bibnamefont{and}
  \bibinfo{author}{\bibfnamefont{D.~A.} \bibnamefont{Huse}},
  \bibinfo{journal}{\prl} \textbf{\bibinfo{volume}{56}}, \bibinfo{pages}{1601}
  (\bibinfo{year}{1986}).

\bibitem[{\citenamefont{Temesv{\'a}ri et~al.}(2002)\citenamefont{Temesv{\'a}ri,
  De~Dominicis, and Pimentel}}]{rscikk}
\bibinfo{author}{\bibfnamefont{T.}~\bibnamefont{Temesv{\'a}ri}},
  \bibinfo{author}{\bibfnamefont{C.}~\bibnamefont{De~Dominicis}},
  \bibnamefont{and} \bibinfo{author}{\bibfnamefont{I.~R.}
  \bibnamefont{Pimentel}}, \bibinfo{journal}{Eur. Phys. J. B}
  \textbf{\bibinfo{volume}{25}}, \bibinfo{pages}{361} (\bibinfo{year}{2002}),
  \eprint{cond-mat/0202162}.

\bibitem[{\citenamefont{Temesv{\'a}ri et~al.}(2000)\citenamefont{Temesv{\'a}ri,
  Kondor, and De~Dominicis}}]{reparametrization}
\bibinfo{author}{\bibfnamefont{T.}~\bibnamefont{Temesv{\'a}ri}},
  \bibinfo{author}{\bibfnamefont{I.}~\bibnamefont{Kondor}}, \bibnamefont{and}
  \bibinfo{author}{\bibfnamefont{C.}~\bibnamefont{De~Dominicis}},
  \bibinfo{journal}{Eur. Phys. J. B} \textbf{\bibinfo{volume}{18}},
  \bibinfo{pages}{493} (\bibinfo{year}{2000}), \eprint{cond-mat/0007340}.

\bibitem[{\citenamefont{De~Dominicis}(2005)}]{Cirano}
\bibinfo{author}{\bibfnamefont{C.}~\bibnamefont{De~Dominicis}}
  (\bibinfo{year}{2005}), \eprint{cond-mat/0509096}.

\bibitem[{\citenamefont{Green et~al.}(1983)\citenamefont{Green, Moore, and
  Bray}}]{GrMoBr83}
\bibinfo{author}{\bibfnamefont{J.~E.} \bibnamefont{Green}},
  \bibinfo{author}{\bibfnamefont{M.~A.} \bibnamefont{Moore}}, \bibnamefont{and}
  \bibinfo{author}{\bibfnamefont{A.~J.} \bibnamefont{Bray}},
  \bibinfo{journal}{J. Phys. C} \textbf{\bibinfo{volume}{16}},
  \bibinfo{pages}{L815} (\bibinfo{year}{1983}).

\bibitem[{\citenamefont{Pimentel et~al.}(2002)\citenamefont{Pimentel,
  Temesv{\'a}ri, and De~Dominicis}}]{Iveta}
\bibinfo{author}{\bibfnamefont{I.~R.} \bibnamefont{Pimentel}},
  \bibinfo{author}{\bibfnamefont{T.}~\bibnamefont{Temesv{\'a}ri}},
  \bibnamefont{and}
  \bibinfo{author}{\bibfnamefont{C.}~\bibnamefont{De~Dominicis}},
  \bibinfo{journal}{\prb} \textbf{\bibinfo{volume}{65}},
  \bibinfo{pages}{224420} (\bibinfo{year}{2002}), \eprint{cond-mat/0204615}.

\bibitem[{\citenamefont{Wegner}(1972)}]{Wegner}
\bibinfo{author}{\bibfnamefont{F.~J.} \bibnamefont{Wegner}},
  \bibinfo{journal}{\prb} \textbf{\bibinfo{volume}{5}}, \bibinfo{pages}{4529}
  (\bibinfo{year}{1972}).

\bibitem[{\citenamefont{Moore}(2005)}]{Moore}
\bibinfo{author}{\bibfnamefont{M.~A.} \bibnamefont{Moore}}
  (\bibinfo{year}{2005}), \eprint{cond-mat/0508087}.

\end{thebibliography}

\end{document}